\documentclass[preprint,12pt]{elsarticle}
\usepackage[utf8]{inputenc}
\usepackage[french, english]{babel}
\usepackage[T1]{fontenc}
\usepackage{amsmath}
\usepackage{amsfonts}
\usepackage{amssymb}
\usepackage{graphicx}
\usepackage[figurename=Fig.]{caption}
\usepackage{subcaption}
\usepackage{thmtools, thm-restate}
\usepackage{hyperref}

\usepackage{amssymb}

\newtheorem{thm}{Theorem}[section]

\newtheorem{proof}{Proof}
\journal{}

\begin{document}

\begin{frontmatter}



\title{Optimal Control applied to SIRD model of COVID 19\tnoteref{title}$^*$}

\author[inst1,inst2]{Amira Bouhali\corref{Bouhali}$^*$}
\ead{amira.bouhali.2893@gmail.com}
\affiliation[inst1]{organization={ENIT},
            addressline={Université de Tunis El Manar}, 
            postcode={1068}, 
            state={Tunis},
            country={Tunisia}}

\author[inst2]{Walid Ben Aribi}
\author[inst2]{Slimane Ben Miled}
\author[inst2,inst3]{Amira Kebir}
\affiliation[inst2]{organization={BIMS},
            addressline={Institut Pasteur de Tunis}, 
            city={ Tunis Belvédère},
            postcode={1002}, 
            state={Tunis},
            country={Tunisia}}
            
\affiliation[inst3]{organization={IPEIT},
            addressline={Université de Tunis}, 
            city={Monfleury},
            postcode={1008}, 
            state={Tunis},
            country={Tunisia}}           

\begin{abstract}
In this study, we present an epidemic controlled SIRD model with two types of control strategies: mask wear and screening. The aim of this study is to minimize the number of the Deceased keeping a minimal cost of mask advertising and screening.  The model is proved to be well posed and to have an invariant region . Also, a thorough study of the dynamics is effected and the basic reproduction number is used to study the stability of the steady states of the model. As for the optimal control analysis, the existence of an optimal control was checked. Then its characterization was carried out using the Pontryagin’s minimum principle . Numerical simulations are conducted after that with different values of maximal screening for comparison. The findings of the optimal control analysis and numerical simulations both reveal that  the optimal pair of strategies contribute enormously in lowering the number of infected and dead individuals.
 Although zero infection is not achieved in the population, this study implies that carrying an optimal approach constitutes a major step in controlling the spread of the disease to the barest minimum that can buy time for middle and low-income countries to carry on with their vaccination strategies.
\end{abstract}


\begin{highlights}
\item Applying mask wear and screening strategies can contain the pandemic.
\item Optimal screening and mask wear allow disease containing with lower cost.
\item High- maximal values of mask wear and  screening provide very efficient and rapid results.
\end{highlights}

\end{frontmatter}


\section{Introduction}
\label{sec:sample1}
Severe acute respiratory syndrome coronavirus 2 commonly known as SARS-CoV-2 is a novel coronavirus that has caused the global pandemic of COVID-19 first reported in Wuhan China a year ago. Soon after, the virus has spread in a very rapid way to extend to the whole world making the World Health Organization declare a global pandemic on March $11^{th}\, , \; 2020$. The virus has proved to be very difficult to contain out of the quarantine measures due to its high contagion that lead to over $180000000$ people to be infected worldwide. However its high contagion is not the only problem as it has proven as well to be lethal  causing over $3900000$ deaths around the world. On the other hand, the economic pressure on the governments has shown how inconvenient the lockdown strategy is on a long term and how much required it is to carry on with a more normal way of life. For that, the problem has been treated not only biologically in the purpose of vaccine and treatment implementation but also mathematically to study its social effects. This is not a first as mathematical modeling has provided a very powerful tool for investigating the dynamics of infectious diseases and controlling them. Previous studies have introduced different models allowing to predict and assess intervention strategies during pandemic spread \citep{ Hassan1, Hassan2} such as Ebola \citep{Harout}, Tuberculosis \citep{Jung} or the current Covid-19 \citep{Abhishek, Ellina}.  Some models assumed life-long immunity such as SIR models, others considered the lower limit as a recovered individual is not supposed to be immune for no matter how short a period and they introduced the SIS models. Some also were more realistic and assumed a gained immunity for a period of time revealing the SIRS models. In our case, we consider an SIRD model where we introduce the disease-caused death equation into the model dynamics as our focus, in the second part of this study, is on minimising the number of these deaths. The optimal control efforts serve mainly that end. Thus, the ultimate goal of this study is to minimize the number of deaths with basic strategies only: mask wear and screening at a  minimal cost. This presents the possibility of containing the disease without any extreme measures such as lockdown or vaccination which represents a very suitable solution for middle and specially low-income countries. As it allows them to minimize the costs and provides time so that they can carry on with their vaccination strategies.
In this work,  both mathematical and numerical analysis of a controlled epidemiological model of four sub-populations: susceptible, infectious, recovered and dead are presented. Section $2$ is a study of the dynamics of the SIR model, its equilibria and their stability. Section $3$ focuses on the optimal control problem that aims to reduce the number of the deceased keeping a minimal screening cost. Section $4$ is dedicated to the numerical simulations and the discussion. Then, a conclusion was drawn in the last section.   
\section{Model Description and analysis}
This section outlines the formulation of a deterministic SIRD model for COVID-19.  The total population at time $t$ is divided into four sub-populations: Susceptible, $S(t)$ ; Infectious, $I(t)$ ; Recovered, $R(t)$ and Dead, $D(t)$.
Two types of control $u_1(t)$ and $u_2(t)$ are used where $1- u_1(t)$ is the probability of mask wear and $u_2(t)$ is the screening rate.
In the Susceptible compartment, $S(t)$,  people are recruited into the population at a constant rate, $\Lambda $, through migration/birth. They exit this compartment either through infection induced by the disease with the force of infection, $u_1 \, \beta\, I(t) $ or natural mortality. The infectious compartment, $I(t)$, gains population through infection induced by the disease at the rate of $ u_1(t) \, \beta\, S(t) $. A proportion, $\alpha $, exits this compartment through recovery at a rate $ u_2(t)+ \delta $ after screening or end of incubation period,  the remaining proportion, $1-\alpha$, of the infectious individuals leaves this compartment at a rate $ u_2(t)+ \delta $ towards the dead compartment through disease induced death, $D(t)$. Recovered individuals are assumed to develop permanent immunity to COVID-19, and compartments, $S\,, \;I$ and $R$ are assumed to have a natural mortality rate,  $\mu$. Therefore, the epidemic model is  given by the following system:
\begin{equation}
\left\{\begin{array}{l}
\frac{dS(t)}{dt} = \mathrm{\Lambda}-u_1(t)\beta S(t)I(t)-\mathrm{\mu}S(t)\\
\frac{dI(t)}{dt} = u_1(t)\beta S(t)I(t)-\left(u_2(t)+\mathrm{\mu}+\right.\mathrm{\delta})I(t)\\
\frac{dR(t)}{dt} =\mathrm{\alpha} \left(u_2(t)+  \mathrm{\delta} \right)I(t)-\mathrm{\mu}R(t)\\
\frac{dD(t)}{dt} =\left(1-\right.\mathrm{\alpha}) \left(u_2(t)+  \mathrm{\delta}\right)I(t)
\end{array} \right. 
\label{2.1}
\end{equation}
subject to the following initial conditions
\begin{equation*}
S(0)\geq 0, I(0)\geq 0, R(0)\geq 0, D(0)\geq 0 
\end{equation*}
All parameters of this model are considered positive .

In what follows, we will study the dynamic of the sub-model, susceptible, infected and recovered ($ S I R $) model, in the case where controls are constants.

\subsection{Analysis of the SIR model with constant controls}

The SIR model corresponds to the first three equations of the system  (\ref{2.1}):

\begin{equation}
   \left\{ \begin{array}{l}
\frac{dS(t)}{dt} = \mathrm{\Lambda}-u_1\beta S(t)I(t)-\mathrm{\mu}S(t) \\
\frac{dI(t)}{dt} = u_1\beta S(t)I(t)-\left(u_2+\mathrm{\mu}+\right.\mathrm{\delta})I(t) \\
\frac{dR(t)}{dt} =\mathrm{\alpha} \left(u_2+  \mathrm{\delta} \right)I(t)-\mathrm{\mu}R(t)
    \end{array}\right.
    \label{3.1}
\end{equation}

We aim here to understand the impact of time independent control parameters,  i.e., $u_1(t)=u_1$ and $u_2(t)=u_2$, on the transmission dynamics of the COVID-19.  

By the following,  we prove that the solutions are uniformly bounded in a positive invariant region,
\begin{equation}
\Omega=\lbrace\left( S, I, R\right)\in\mathbb{R}_+^3:  S + I + R \leq \frac{\Lambda}{\mu} \rbrace \label{3.3}
\end{equation}

\begin{thm}
For any non-negative initial condition, the solution of system \eqref{3.1} remains non-negative and positively bounded. In addition, the set $ \Omega $ is positively invariant for the epidemic model \eqref{3.1}.
\end{thm}

\begin{proof}
 The positivity of the solutions of the system \eqref{3.1} can be verified by examining the direction of the vector field $ \left(\frac{dS(t)}{dt}, \frac{dI(t)}{dt}, \frac{dR(t)}{dt} \right)^T  $ of \eqref{3.1} on each coordinate plane.
In the $ \lbrace IR \rbrace $ hyper-plane, for $ S = 0 $, one has, 
$$\frac{dS(t)}{dt} \mid_{S = 0} = \Lambda \geq 0.$$
This shows that the vector field points to the interior of $\mathbb{R}_+^3 $. Therefore, no trajectory can leave the positive octant by crossing the boundary face $S=0 $.\\
Solutions starting from the $ \lbrace IR \rbrace $ hyper plane remain in the same hyper plane.\\
Similarly, if $ R=0 $ and $ I > 0 $, one has 
 $ \frac{dR(t)}{dt}\mid_{R = 0} =  \mathrm{\alpha} \left(u_2+  \mathrm{\delta} \right)I(t) > 0.$
Thus, no trajectory can leave through the boundary face $ R=0 $.\\
Likewise, if $ I = 0$ and $ R > 0$ one has $ \frac{dI(t)}{dt}\mid _{I = 0} =  0 .$
This indicates  that once a  trajectory enters this boundary face, it will remain there.  
In addition, we have:
\begin{align*}
\frac{d I(t)}{dt}\mid _{I = 0} &=  0 \\ 
\frac{d R(t)}{dt}\mid _{R = 0} &=  \mathrm{\alpha} \left(u_2+  \mathrm{\delta} \right)I(t) > 0
\end{align*}

In addition, let  $N (t) = S (t) + I(t) + R(t) $ be the total population number at time $t$, then $\forall t\in{R}_+$:
$$\begin{array}{l c l}
\frac{dN}{dt}\left(t\right) & = & \Lambda - \mu \left(S(t)+I(t)+R(t)\right) -\left(1-\alpha)(u_2+\delta \right)I(t) \\
     &\leq & \Lambda - \mu \left(S(t)+I(t) +R(t)\right) \\
      &\leq & \Lambda - \mu N\\
      &\leq &- \mu \left( N(t) - \frac{\Lambda}{\mu}\right)\\
\end{array}$$
Consequently, according to Gronwall's lemma, one has
$$ N(t) - \frac{\Lambda}{\mu}  \leq \left( N(0) - \frac{\Lambda}{\mu} \right) e^{-\mu t} $$
 
 and then, 
$$ N(t) \leq \frac{\Lambda}{\mu} + \left( N(0) - \frac{\Lambda}{\mu}\right)e^{-\mu t} \leq max (N(0), \frac{\Lambda}{\mu}) = \frac{\Lambda}{\mu} $$

\end{proof}

\subsubsection*{  Existence and global stability of equilibrium points
}\label{sec:nothing2}
In this section the existence and the stability of disease-free equilibrium and the endemic  equilibrium states of model \eqref{3.1} are examined. 

First, we need to define the basic reproduction number, $R_0$. This quantity predicts the spread of a disease in the population. It is defined as the  average number of secondary infections generated when an infected person is introduced into a host population where everyone is susceptible and it is given by :
\begin{equation}
R_0=  \frac{\partial_I F(S, I, R)}{\partial_I V(S, I, R)}|_{ ( \frac{\Lambda}{\mu}, 0, 0) }=\frac{u_1\beta \Lambda}{\mu \left( u_2+\mu+\right.\delta)} 
\end{equation}
where $  F(S, I, R) = u_1(t)\beta S(t)I(t) $ and $ V (S, I, R)= \left(u_2(t)+\mathrm{\mu}+\right.\mathrm{\delta})I(t)$ denote respectively the rates of the transfer in and out of the infected compartment.

Then, It is easy to show that the system \eqref{3.1} has two steady states: a disease-free equilibrium (DFE)  given by $ E_0^* =  (\frac{\Lambda}{\mu}, 0, 0 )$ that exists for any value of the parameters and an endemic equilibrium $ E_1^* = (S^*, I^*, R^* ) $ in the interior of $ \Omega $ that exists if and only if $R_0 > 1$  and where, 
$$S^* = \frac{\Lambda}{\mu R_0}, I^* =  \frac{\Lambda}{u_2+\mu+\delta}\left[ 1-\frac{1}{R_0}\right],  R^* =\frac{\alpha (u_2+\delta) (R_0 - 1)}{u_1 \beta}. $$

For the global stability of  equilibrium we use popular types of Lyapunov functions i.e, the common quadratic and Volterra-type functions.

\begin{thm}
If $ R_0\leq 1 $, then the DFE, $E_0^* $, is globally asymptotically stable on $ \Omega $. If $ R_0>1 $, then the endemic equilibrium, $E_1^* $, is globally asymptotically stable.
\end{thm}

\begin{proof}
Since $R$ is not present in the first two equations of the system \eqref{3.1}, then by  theorem 3.1 of \citep{13}, to study the stability, it is sufficient to analyze, the following isolated subsystems:

\begin{equation}
   \left\{ \begin{array}{l}
\dot{S} = \mathrm{\Lambda}-u_1\beta SI-\mathrm{\mu}S \\
\dot{I} = u_1\beta SI-\left(u_2+\mathrm{\mu}+\right.\mathrm{\delta})I
 \end{array}\right.
    \label{3'.1}
\end{equation}

and 

\begin{equation}
 \begin{array}{l}
\dot{R} =\mathrm{\alpha} \left(u_2+  \mathrm{\delta} \right)I-\mathrm{\mu}R
    \end{array}
   \label{3'.2}
\end{equation}

It is obvious that  equation \eqref{3'.2}  is globally exponentially stable for  $I=0$ or $I=I^*$ .

As for the global asymptotic stability of \eqref{3'.1}, one can  use the Lyapunov function and LaSalle's theorem \cite{LaSalle1, LaSalle2}.
  
For the DFE, we define by $ \Omega_1= \{(S, I) \in \mathbb{R}_+^2: S+I\leq \frac{\Lambda }{  \mu }\}$ and  

${V}: \Omega_1 \rightarrow \mathbb{R}$ by
$${V}(S, I)=\frac{1}{(u_2+\mu+\delta)}  I $$
Then, if $R_{0} \leqslant 1$ one has
$$
\begin{array}{llll}
\dot{V}(S,I) & = \frac{1}{u_2+\mu+\delta}\dot{I} &=(\frac{1}{u_2+\mu+\delta} u_1 \beta SI-I) &=(\frac{ \mu }{\Lambda  }R_0S-1)I\\
            &\leq(R_0-1)I & & \\
           & \leq 0 & & 
\end{array}
$$

If $\dot{V}=0$ then $I=0$ or $S=\frac{\Lambda}{\mu}$ and $R_{0}=1 $. Then the the largest compact invariant set in $\{(S,I) \in \Omega_1, \dot{V}(S,I) = 0\}$ is the singleton $\{(\frac{\Lambda }{  \mu }, 0)\}$. It follows from the LaSalle's invariance principle \cite{LaSalle2} that $(\frac{\Lambda }{  \mu }, 0)$ is globally asymptotical stable for \eqref{3'.1}. Therefore, using   Theorem3.1 of \citep{13}, we conclude that $E_0^*$ is an asymptotically stable equilibrium point of \eqref{3.1}.

If now $R_0>1$,  the endemic equilibrium exists, and then we define the Lyapunov function ${L}: \{(S,I)\in\Omega_1: S>0,I>0\} \rightarrow \mathbb{R}$ by:
\begin{align}
L(S,I, R) = S - S^* - S^* \ln \left(\frac{S}{S^*}\right)+ I - I^*- I^* \ln \left(\frac{I}{I^*}\right)
\end{align}

 $L$ is $C^1$ on the interior of $\Omega_1$, $E^*$ is the global minimum of $L$ on $\Omega_1$, and $L(S^* ,I^*)= 0$.

The time derivative of $L$ computed along solutions of \eqref{3.1} is
{\small 
$$ \begin{array}{rl}
\dot{L}(S,I,R) & =  \dot{S} - \frac{S^*}{S} \dot{S} + \dot{I} - \frac{I^*}{I} \dot{I} \\
  & =\Lambda - u_1 \beta SI-\mu S-\frac{S^*}{S}\left( \Lambda - u_1 \beta SI-\mu S\right) + u_1 \beta SI-\left(u_2+\mu+\right.\delta)I\\
  & - \frac{I^*}{I}\left(u_1 \beta SI-\left(u_2+\mu+\right.\delta)I \right) \\
 & = \Lambda -\mu S-\frac{S^*}{S}\Lambda+ u_1 \beta S^* I+\mu S^* -\left(u_2+\mu+\right.\delta)I -  u_1 \beta S I^*+\left(u_2+\mu+\right.\delta)I^* 
\end{array}$$}

Knowing that, $ \Lambda = u_1 \beta S^* I^*+\mu S^* $, then we have

\begin{align*}
\dot{L}(S,I,R) &=  u_1 \beta S^{*}I^{*}+\mu S^{*} - \mu S - \frac{S^*}{S}\left[u_1 \beta S^{*}I^{*}+\mu S^{*} \right]\\ 
 &+ u_1 \beta S^{*}I+\mu S^* - \left(u_2+\mu+\right.\delta)I - u_1 \beta SI^{*}+\left(u_2+\mu+\right.\delta)I^* \\
 &= u_1 \beta S^{*}I^{*} +\mu S^{*} - \frac{S}{S^*}\mu S^* -\frac{S^*}{S}u_1 \beta S^{*}I^{*}-\frac{S^*}{S}\mu S^{*} \\
 &+ \frac{I}{I^*}\left[u_1 \beta S^{*}I^* - \left(u_2+\mu+\right.\delta)I^* \right] +\mu S^* - \frac{S^*}{S} u_1 \beta S^*I^{*}+u_1 \beta S^{*}I^{*} 
\end{align*}

Using that $  u_1 \beta S^* I^* - \left(u_2+\mu+\right.\delta)I^*=0 $,

then 

\begin{align*}
\dot{L}(S,I,R) &=  (u_1 \beta S^* I^* + \mu S^*)\left[2-\frac{S^*}{S}- \frac{S}{S^*} \right]  \\
&=  -\Lambda\frac{1}{SS^*}(S- S^*)^2 \\
&\leq 0  
\end{align*}

Then $\dot{L}$ is negative definite and $\dot{L}=0$ if and only if $S=S^{*}$. Therefore the largest compact invariant set in $\left\{(S, I) \in \Omega_1: \dot{L}=0\right\}$ is the singleton $\left\{(S^{*},I^{*})\right\}$.  Indeed if $S=S^*$ we get $\dot{S}=0$ and then $I=I^*$. By LaSalle's invariance principle, we conclude that $(S^{*},I^{*})$ is globally asymptotically stable for the system \eqref{3'.1}. Therefore, using   Theorem3.1 of \citep{13}, we conclude that $E_1^*$ is an asymptotically stable equilibrium point of \eqref{3.1}.
\end{proof}

\subsection*{Effect of constant mask wear and screening}
\begin{figure}[!h]
    \centering
    \includegraphics[scale=0.5]{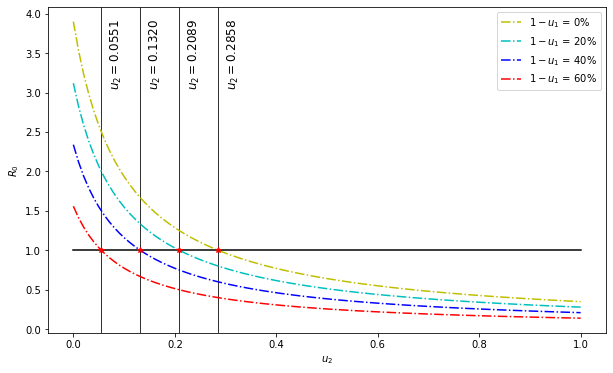}
    \caption{The reproduction rate vs screening rate for different values of maximal mask wear $1- u_1$ }
    \label{fig:1}
\end{figure}

In this section, we investigate the effect of mask wear and screening on the spread of the disease. The two strategies are still assumed constant.  First it is clear that for a $100\% $ of mask wear i.e. $u_1 = 0$,  one has $R_0 = 0$ and thus there is no transmission of the disease. Screening at this point is not needed . On the other hand, for $ 0\% $ of mask wear i.e. $u_1 = 1$ one has $R_0 = \frac{\beta \Lambda}{\mu \left( u_2+\mu+\right.\delta)}  $. And since stopping the spread of the disease requires that $ R_0 < 1 $, it becomes necessary for the screening value to exceed $ \frac{\Lambda\, \beta}{\mu} - (\delta + \mu) $ in order to keep the spread of the disease under control. To investigate the effect of a combination of mask wear and screening,  $R_0$ is plotted for different values of mask wear $1-u_1 \in \{ 0;~ 0.2;~ 0.4;~ 0.6;~ 0.9\} $. The curve of $R_0$ is decreasing and it crosses the value of one at some point.  We noticed that the more people wear masks, the less screening is needed and the faster $R_0$ reaches the value $1$, see figure (\ref{fig:1}).

\begin{figure}[!h]
   \begin{subfigure}{.5\textwidth}
       \centering
       \includegraphics[scale=0.5]{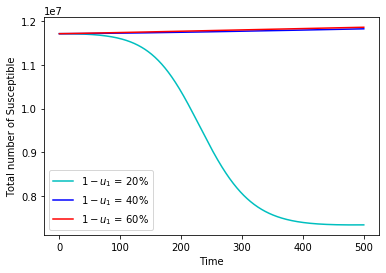} 
       \label{fig:sub-second}
   \end{subfigure}
   \begin{subfigure}{.5\textwidth}
       \centering
       \includegraphics[scale=0.5]{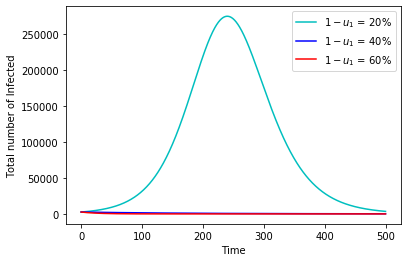}
       \label{fig:sub-first}
   \end{subfigure}
\caption{Total number of susceptible  and infected  for three different values of mask wear $ 1- u_1 $ : red $1-u_1 = 0.6$, blue $1- u_1 = 0.4$ and cyan $1-u_1= 0.2$ and for $u_2=0$ }
\label{fig:2}
\end{figure}

Hence, it is obvious that  wearing masks, even without screening,  helps in flattening the infection curve and increasing the number of susceptible, see figure (\ref{fig:2}).

\newpage
\section{Optimal control}
As optimal control techniques are of great use in developing optimal strategies to control various kinds of diseases we use them in this section with the aim of reducing the number of deceased individuals at a finite time, $D(t^f)$, keeping a minimal cost of screening $\int_0^{t^f}u_2^2(t) dt$.  \\
Therefore, the objective function that we seek to minimize over a finite time horizon $[0,t^f]$ is given by:
\begin{equation}
\begin{array}{lclcl}
J(u_1,u_2) &=& A_1D(t^f)+A_2\int_0^{t^f}u_2^2(t)dt \\
&=& \int_0^{t^f} A_1 \left(1-\right.\mathrm{\alpha})\left(u_2(t)+\right.\mathrm{\delta})I(t) +  A_2 u_2^2(t)dt
\end{array}
\label{cost}
\end{equation}

Where the set of admissible controls $U$ is given by

$$ U = \{ u =(u_1(t), u_2(t))  \in \left( L^{\infty}( 0; t_f)\right)^2 \mid 0\leq u_i^{min}\leq u_i(t) \leq u_{i}^{max}\leq 1\, , \mathrm{for} \; i = 1, 2  \} $$
Note that the controls are no longer considered constant.
\begin{thm}
There exists an optimal control $ u^* $ and a corresponding state variables vector $ \left( S^0, I^0, R^0, D^0\right)$ that minimizes the objective function.
\end{thm}
\begin{proof} 
The existence of the optimal control pair can be obtained using a result by Fleming and Rishel (1975) and by Lukes (1982) \cite{Hassan1,book2}. \\ 
In fact, one can easily verify that:\\
1. The set of controls and corresponding state variables is nonempty.\\
2. The admissible set $ U $ is convex and closed.\\
3. The right hand side of the state system \ref{2.1} is bounded by a linear function in the
state and control variables.\\
4. The integrand of the objective functional $L$ is convex on $U$ and there exists constants $\omega_1 >0 , \omega_2 >0  $ and $ \rho >1$ such that 
$$ L(u) \geq \omega_2 + \omega_1 ( |u_1|^2 + |u_2|^2)^{\frac{\rho}{2}} .$$
\end{proof}
In order to determine the optimal control,  Pontryagin's Minimum Principle was used \citep{book2}. The latter changes the optimality system into a study of the Hamiltonian variations through the use of adjoint functions . The Hamiltonian is given by 

$$ H(t, u, X, \lambda) = <\lambda(t), \dot{X}(t)> + A_1  \left(1-\mathrm{\alpha}\right)\left(u_2(t)+\mathrm{\delta}\right)I(t)+ A_2 u_2^2 (t) $$
where  $ X = (S, I, R, D) $ is the vector of state variables and $ \lambda = (\lambda_1(t), \lambda_2(t), \lambda_3(t), \lambda_4(t)) $ is the vector of adjoint variables and $<.,.> $ is the scalar product. 

\begin{thm}
Given optimal controls $u^*_1(t) \, , \; u^*_2(t) $ and the corresponding solution $S^0(t)\, ,\;  I^0(t) \, ,\;  R^0(t) $ and $D^0(t)$ of the corresponding state system (\ref{2.1}) - (\ref{cost}), there exists adjoint variables $\lambda_1 \, , \;\lambda_2\, , \; \lambda_3$ and $\lambda_4$ that satisfy
\begin{equation}
\begin{array}{lcl}
  \dot{\lambda}_1 & = &  \beta I (\lambda_1 - \lambda_2) u_1 +\lambda_1 \mu \\
\dot{\lambda}_2& = &  \beta S (\lambda_1 - \lambda_2) u_1 +\lambda_2( u_2 + \mu + \delta) - A_1 ( 1- \alpha) (u_2 + \delta)   \\
\dot{\lambda}_3& = &  \mu \lambda_3  \\
\dot{\lambda}_4& = &   0
\end{array}
\label{12}
\end{equation} 
  with transversality conditions: 
  \begin{equation}
     \lambda_i (t_f) = 0 \, , \quad i = 1,\,2,\,3,\,4. 
  \end{equation}

 Furthermore, the optimal control is given by $u^* = (u^*_1,u^*_2) $ where
$$ u^*_1 = \left\lbrace
\begin{array}{l l}
u_1^{min}  \; & , \mathrm{if }\, \lambda_2 - \lambda_1 > 0\\
u_1^{max} \; & , \mathrm{if }\, \lambda_2 - \lambda_1 <0\\ 
\end{array} 
\right.$$ 
$$ u^*_2 = \left\lbrace
\begin{array}{l l}
\frac{\left( \lambda_2 - ( 1 -\alpha)A_1 \right)I}{2 A_2} \; & , \mathrm{if }\, u_2^{min} < \frac{\left( \lambda_2 - ( 1 -\alpha)A_1 \right)I}{2 A_2} < u_2^{max}  \\
u_2^{min}        \; & , \mathrm{if }\, \frac{\left( \lambda_2 - ( 1 -\alpha)A_1 \right)I}{2 A_2} < u_2^{min}  \\
u_2^{max} \; & , \mathrm{if }\,\frac{\left( \lambda_2 - ( 1 -\alpha)A_1 \right)I}{2 A_2} > u_2^{max} \\ 
\end{array} \right. $$
\end{thm}

\begin{proof}
According to Pontryagin's minimum principle,
$$\dot{\lambda} = - \frac{\partial H}{\partial X} .$$ 
Thus, the adjoint functions $ (\lambda_1(t), \lambda_2(t), \lambda_3(t), \lambda_4(t)) $ have the following dynamics
\begin{equation*}
\begin{array}{lcl}
  \dot{\lambda}_1 & = &  \beta I (\lambda_1 - \lambda_2) u_1 +\lambda_1 \mu  \\
\dot{\lambda}_2 & = & \beta S (\lambda_1 - \lambda_2) u_1 +\lambda_2( u_2 + \mu + \delta) - \lambda_3 \alpha  (u_2 + \delta)- ( 1- \alpha) (u_2 + \delta) \lambda_4 - A_1 ( 1- \alpha) (u_2 + \delta)   \\
\dot{\lambda}_3 & = &  \mu \lambda_3  \\
\dot{\lambda}_4 & = &   0    \\
\end{array}
\end{equation*} 
with the final conditions 
$$ \lambda(t^f) = (0, 0, 0, 0). \label{4.2.5} $$ 
From  the third and fourth equations  we can deduce that $ \lambda_3 \equiv 0 $ and $ \lambda_4 \equiv 0 $.\\
Consequently, the Hamiltonian becomes 
\begin{equation}
\begin{array}{lcl}
H &= &\left(\mathrm{\Lambda}-u_1(t) \beta S(t)I(t)-\mathrm{\mu}S(t)\right) \lambda_1 +\left(  u_1(t)\beta S(t)I(t)-\left(u_2(t)+\mathrm{\mu}+\right.\mathrm{\delta})I(t) \right) \lambda_2 \\
  &  & + A_1 \left(1\right.\mathrm{\alpha})\left(u_2(t)+\right.\mathrm{\delta})I(t)+ A_2 u_2^2 (t). 
\end{array}
\end{equation}
and the adjoint variables dynamics is reduced to 
$$ \begin{array}{lcl}
  \dot{\lambda}_1 & = &  \beta I (\lambda_1 - \lambda_2) u_1 +\lambda_1 \mu \\
\dot{\lambda}_2& = &  \beta S (\lambda_1 - \lambda_2) u_1 +\lambda_2( u_2 + \mu + \delta) - A_1 ( 1- \alpha) (u_2 + \delta)   \\
\dot{\lambda}_3& = &  \mu \lambda_3  \\
\dot{\lambda}_4& = &   0
\end{array}
\label{12} $$
Also, the Pontryagin's Minimum Principle states that the optimal control $ u^* $ minimizes the Hamiltonian, hence we should seek the minimum of $ H $ . So we need to study the critical points of the Hamiltonian.
A critical point of $ H $, $u^*= (u^*_1, u^*_2)  $ satisfies $\frac{d\, H}{d \, u} = 0 $ where 
$$
\left\lbrace
\begin{array}{lcl}
\frac{\partial H}{\partial u_1}& = & \beta S I ( \lambda_2 - \lambda_1)  \\
\frac{\partial H}{\partial u_2}& = &  \left(- \lambda_2 + ( 1 -\alpha)A_1 \right)I  + 2 A_2 u_2  \\
\end{array}
\right. $$
The equation $\frac{\partial H}{\partial u_2} = 0 $ implies that 
$$ u^*_2 = \frac{\left( \lambda_2 - ( 1 -\alpha)A_1 \right)I}{2 A_2} $$
The first equation however, shows that the minimum is either reached at $ u^*_1 = u_1^{min} $ or  $ u^*_1 = u_1^{max} $ according to the sign of $ \lambda_2 - \lambda_1 $. \\
In fact when $ u_1 $ is supposed constant;  $ H $ would depend on $ u_2 $ only and  therefore $ u^*_2 $ is a minimum to $ H $ since $ A_2 > 0 $. In that case, one has
$$ H(u_1, u_2) > H (u_1,u^*_2) $$
Since $ u_1^{min}  \leq u_1 \leq u_1^{max} $ then two scenarios are possible
\begin{itemize}
\item If $ \beta S I( \lambda_2 - \lambda_1) > 0 $ i.e. $ \lambda_2 - \lambda_1 > 0 $ then 
$$(\lambda_2 - \lambda_1) u_1^{min}  \leq (\lambda_2 - \lambda_1) u_1 \leq (\lambda_2 - \lambda_1) u_1^{max}  $$
and consequently, 
$$ H(u_1, u_2) \geq H (u_1, u^*_2) \geq H (u_1^{min},  u^*_2)$$
\item If $ \beta S I ( \lambda_2 - \lambda_1) < 0 $ i.e. $ \lambda_2 - \lambda_1 < 0 $ then 
$$ (\lambda_2 - \lambda_1) u_1^{min}  \geq (\lambda_2 - \lambda_1) u_1 \geq (\lambda_2 - \lambda_1) u_1^{max}  $$
and consequently, 
$$ H(u_1, u_2) \geq H (u_1,  u^*_2) \geq H (u_1^{max},  u^*_2)$$
\end{itemize}
Note that $ u^*_2 $ must satisfy $ u_2^{min} <  u^*_2 < u_2^{max} $ to be taken into consideration.
Otherwise,
\begin{itemize}
    \item $\underset{u_2 \in [ u_2^{min}, u_2^{max} ]}{min } H = H(u_2^{min}) \; \mathrm{if} \; \frac{\partial H}{\partial u_2} > 0 \; \mathrm{ i.e.} \; \left(- \lambda_2 + ( 1 -\alpha)A_1 \right)I  + 2 A_2 u_2 > 0$ 
    \item $\underset{u_2 \in [u_2^{min}, u_2^{max}] }{min } H = H(u_2^{max} )\; \mathrm{if} \; \frac{\partial H}{\partial u_2} < 0 \; \mathrm{ i.e.} \;\left(- \lambda_2 + ( 1 -\alpha)A_1 \right)I  + 2 A_2 u_2 <  0$.
\end{itemize}
Assume now that there exists a subset $ [t_0, t_1] \in [0, t_f] $ such that $ \frac{\partial H}{\partial u} = 0 $ for all $ t \in  [t_0, t_1] $. This implies that $$ \left \lbrace 
\begin{array}{lcl}
\beta S I ( \lambda_2 - \lambda_1)& = & 0  \\
\left(- \lambda_2 + ( 1 -\alpha)A_1 \right)I  + 2 A_2 u_2 & = & 0 \\
\end{array}
\right.$$
And consequently
$$ \left \lbrace 
\begin{array}{lcl}
\beta S I( \lambda_2 - \lambda_1)& = & 0  \\
\left(- \lambda_2 + ( 1 -\alpha)A_1 \right)I  & = & 0 \\
 A_2 & = & 0\\
\end{array}
\right.$$
Since $ A_2 > 0 $, we deduce that it is not possible to have $\frac{\partial H}{\partial u_2} = 0 $ and therefore we cannot discuss the case of singular control in the usual terms. However, it is possible to have $\frac{\partial H}{\partial u_1} = 0$ which implies that $\beta S I ( \lambda_2 - \lambda_1) = 0$. Consequently, either $ S . I = 0 $ or $ \lambda_2 - \lambda_1 = 0 $. As the first case does not present quite an interesting case of study, we move to the latter that yields
{\footnotesize
$$\begin{array}{crcl}
    & \lambda_1 & = & \lambda_2 \\
 \Rightarrow & \dot{\lambda_1}  & = & \dot{\lambda_2}\\
 \Rightarrow & \beta I (\lambda_1 - \lambda_2) u_1 +\lambda_1 \mu  & = & \beta S (\lambda_1 - \lambda_2) u_1 +\lambda_2( u_2 + \mu + \delta) - A_1 ( 1- \alpha) (u_2 + \delta) \\
\Rightarrow & ( u_2 +\delta) (\lambda_1 - A_1 ( 1- \alpha) )  & = & 0 \\
\end{array}$$ }
Thus, either $ u_2 = - \delta $ which is not taken into account since  $  - \delta \notin [0; u_2^*] $ or $ \lambda_1 = A_1 ( 1- \alpha) = \lambda_2 $.
However, according to the co-state variables dynamics, one has $ \dot{\lambda}_1 = \mu \lambda_1$ which implies that $ \lambda_1(t) = \lambda_1(t_0) e^{\mu (t-t_0)} $.Consequently, $ \lambda_1(t_0) e^{\mu (t-t_0)} = A_1 ( 1- \alpha) \; ,  \; \forall t \in [t_0, t_1] $. 
This equality is absurd except for one particular case $ \alpha = 1 $ and $  \lambda_1(t_0) = 0 $. Therefore, the existence of  an interval $ [t_0; t_1]  $ such that $ \frac{ 
\partial H}{\partial u} = 0 \;  \forall t \in [t_0; t_1]  $ is not possible.
 \end{proof}

\section{Numerical simulations and discussion}

In this section,  the system (\ref{2.1}) is solved numerically, and  the results obtained  are presented below.
The numerical simulations were carried out by implementing a 4th order Runge-Kutta Method (see, for example \cite{book2}). This iterative method consists in solving the system of equation (\ref{2.1}). Details of the application of this method are developed in \cite{12}.
Then, the adjoint variables equations  are solved by a reverse fourth order Runge-Kutta scheme using the current iteration solution of equation (\ref{12}). The iteration stops if the values of the unknowns at the previous iteration are very close to those of the current iteration.
The parameters used are presented in the table \ref{tab2}.

\begin{center}
{\scriptsize
 \begin{tabular}{c l l l}
\hline
 Parameters    &  Description     &  Values & References\\
 \hline 
$\alpha$  & The rate at which infected individuals become cured    & $\approx 0.99$ & \cite{10} \\
     $N(0)$   & The total size of the population & $11172177 $ &  \cite{9}\\
  $\beta$  & The disease transmission coefficient &  $ 0.15473652/N(0) $ & Fitted \\ 
  $1/\delta$ & The mean duration of infection &  $10.14$ days & Fitted \\
   $\mu$ & The death rate & $0.000017534$  & \cite{9}\\
   $\Lambda$ & The birth rate &  $510.5937$ & \cite{9} \\
   $ A_1$  &  The balancing factor associated to the cost component   &  $30$ &  Assumed \\
    $ A_2$  &  The balancing factor associated to the cost component   &  $10$ &  Assumed \\
     $ 1 - u_1 $  &  Mask wear rate per unit of time    &  $ 0.4 < u_1 < 1$ &  Assumed \\
      $ u_2 $  &  Screening rate per unit of time   &  $ 0 < u_2 < 0.5 $ &  Assumed \\
 \end{tabular} 
 \captionof{table}{Description and values of the parameters}
\label{tab2}}
\end{center}
To start, the system is solved  using the set of parameters listed above and the following initial conditions 
$$[ S(0) = 11718548 ; ~I(0) = 2629 ;~ R(0) = 0 ;~ D(0) = 0]$$

\begin{figure}[!h]
       \centering
       \includegraphics[scale=0.6]{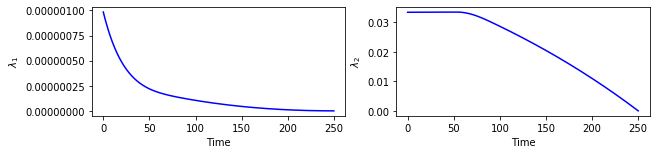}
       \caption{Adjoint variables $\lambda_1$ and $\lambda_2$}
       \label{fig:3}
\end{figure}

\begin{figure}[h!]
       \centering
       \includegraphics[scale=0.6]{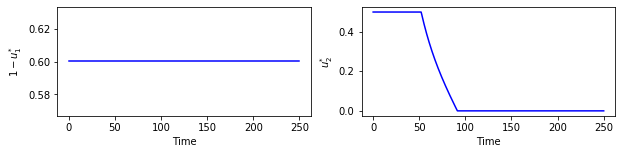}
       \caption{Optimal mask wear rate $1-u_1^*$ (left) and screening rate $u_2^*$ (right) per unit of time}
       \label{fig:sub-second}
\label{fig:3'}
\end{figure}

We introduced the control and solved the optimality system. With the use of these parameters, and the adjoint variables dynamics,  the following solutions for $\lambda_1$ and $\lambda_2$ were obtained. For this set of parameters,  $\lambda_2 - \lambda_1  $ is always positive (see figure \ref{fig:3}). According to the optimal control study conducted above, this results in 
$$ u_1^* = u_1^{min} .$$
For that value of $u_1$, one has maximal constant mask wear while the screening rate starting at a value near $0.5$ remains constant during the first 50 days then starts decreasing until it reaches $0$ (see figure \ref{fig:3'}). 

\begin{figure}[!h]
    \begin{subfigure}{.5\textwidth}
       \centering
       \includegraphics[scale=0.6]{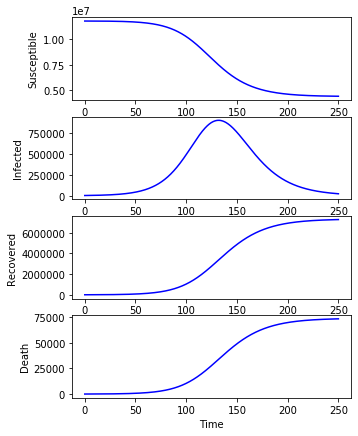}
       \caption{}
       \label{fig:sub-second}
   \end{subfigure}   
   \begin{subfigure}{.5\textwidth}
       \centering
       \includegraphics[scale=0.6]{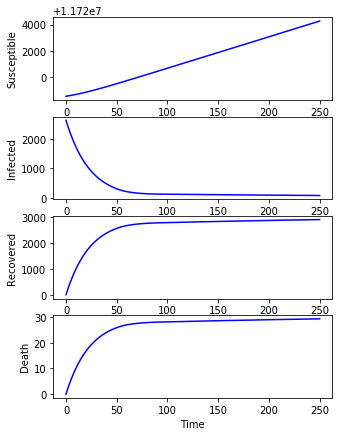}
       \caption{}
       \label{fig:sub-second}
   \end{subfigure}
   
   \caption{ Dynamics of state variables per unit of time in two cases: without any control measures $1-u_1 = u_2 =0$ (a) and with optimal control pair $(u_1^*,u_2^*)$ (b)}
\label{fig:5}
\end{figure}

Then, the state variables were plotted in two cases: controlled and uncontrolled. In the absence of any form of control, the susceptible curve  starts decreasing  at the $50^{th}$ day until it reaches a value near zero. On  the other hand, the curve of the infected reaches a peak that exceeds $ 2 . \, 10^6$ and the number of deaths reaches $3.\, 10^5$. However, once the system is controlled, a huge difference in the dynamics is observed. The susceptible number is increasing as opposed to the the infections  that start at a maximal value of $10^3$ then decrease to a value near 100. The dead curve is still increasing; however, to a maximal value less than $50$ ( see figure \ref{fig:5}).

\begin{figure}[h!]
       \centering
       \includegraphics[scale=0.5]{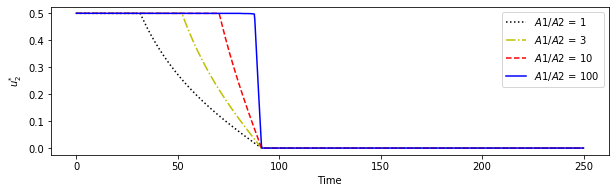}
       \caption{The screening rate $u_2^*$ per unit of time}
       \label{fig:sub-second}
\label{fig:U}
\end{figure}

The coefficients, $A_1$ and $A_2$, are balancing cost factors due to the size and importance of the two parts of the objective function. We assume that the coefficient $A_1$ associated with $D(t^f)$ is greater than or equal to $A_2$, which is associated with the control $u_2$. This assumption is based on the following facts: The cost associated with $D(t^f)$ will include the cost of the dead person, and the cost associated with $u_2$ will include the cost of screening.The fractions of the weighting factors, $A_1/A_2 = 1,~ 3,~ 10$ and $100$, are presented in Figure (\ref{fig:U}).  And to illustrate  the optimal strategy we have chosen the weighing factor, $A1/A2 = 3$ since the only change observed was in the values of the controls rather than the dynamics. 

\begin{figure}[!h]
        \begin{subfigure}{.5\textwidth}
       \centering
       \includegraphics[scale=0.5]{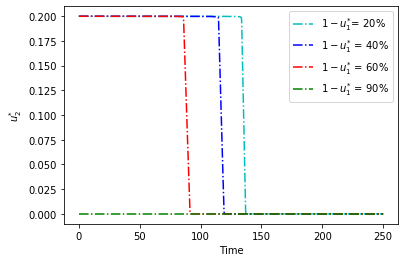}
       \caption{$u_2^*$ with $u_2^{max} = 0.2$}
       \label{fig:sub-second}
   \end{subfigure}
   \begin{subfigure}{.5\textwidth}
       \centering
       \includegraphics[scale=0.5]{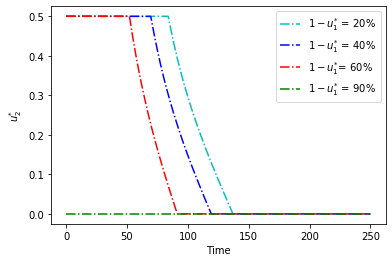}
       \caption{ $u_2^*$ with $u_2^{max} = 0.5$}
       \label{fig:sub-second}
   \end{subfigure}
   \caption{ Optimal screening $u_2^*$ for two different values of upper bound $u_2^{max} \in \{0.2; 0.5\}$ and four different values of maximal mask wear $1-u_1^{min} = 1-u_1^* \in \{ 0.2, 0.4, 0.6, 0.90\}$ }
\label{fig:6}
\end{figure}

 This illustrates the importance of any form of control applied to the population. And since $u_1$ is given by one of the extreme values, it is important to check the effect of the maximal screening and mask wear on the dynamics of the population.
Thus, $u_2^*$ is calculated. For the sake of comparison $u_2$ is also calculated and plotted for several values of $u_1$ as illustrated in figure \ref{fig:6}. 

\begin{figure}[!h]
   \begin{subfigure}{.5\textwidth}
       \centering
       \includegraphics[scale=0.55]{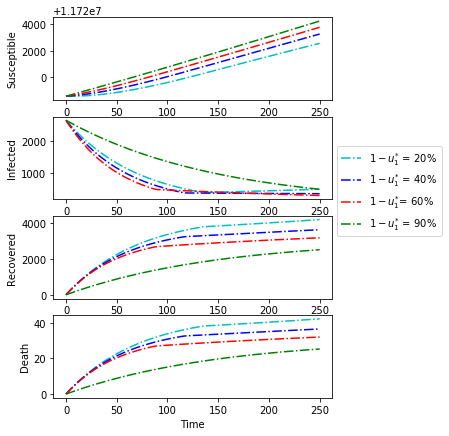}  
       \caption{ $ u_2^{max} = 0.2 $}
       \label{fig:sub-second}
   \end{subfigure}
   \begin{subfigure}{.5\textwidth}
        \centering
        \includegraphics[scale=0.55]{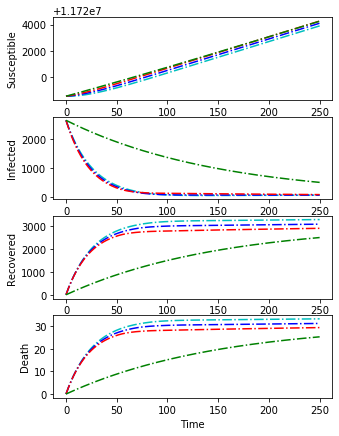} 
        \caption{ $ u_2^{max} = 0.5 $}
        \label{fig:sub-first}
   \end{subfigure}
\caption{Simulation of SIRD model for two values of maximal screening $u_2^{max} \in \{ 0.2, 0.5\} $ and four different values of mask wear $1- u_1^{min} \in \{ 0.2, 0.4, 0.6, 0.90\}$}
\label{fig:7}
\end{figure}
In order to present the importance of maximal mask wear and screening values, the state variables were represented for two  values of maximal screening $ u_2^{max} \in \{ 0.2 ; \,0.5  \} $ and four different values of mask wear $0.2, \,  0.4, \,  0.6, \,  0.9 $ (See figure \ref{fig:7}). 
In the first case, with maximal screening at $0.2$, and for all values of mask wear, the susceptible and recovered were increasing and the dead didn't exceed $60$. However, the infected curve soon showed an increase after having decreased for low mask wear $ 1- u_1 \in \{0.2 , \, 0.4\}$. For $40\%$ of mask wear, the number of infections decreases to $400$ and remains constant after day $50$.
For  $90 \%$ of mask wear, the infected keeps decreasing.  
On the other hand, with $ u_2^{max}= 0.5$, the susceptible and recovered curves reach their maximal values much faster. The dead curve has a maximal value lower than the first case. And interestingly,  the curve of the infected does not show the increase registered in the first case. It also stabilizes at a very low value that is almost equal to zero.

It is important to mention that the values used for the transmission rate $\beta$ and the mean duration of infection $\frac{1}{\delta} $ are not assumed but are rather estimated based on the real values registered by the Tunisian authorities as presented bellow.
\subsection*{Parameter estimation}
\begin{figure}[!h]
    \begin{subfigure}{.5\textwidth}
       \centering
       \includegraphics[scale=0.3]{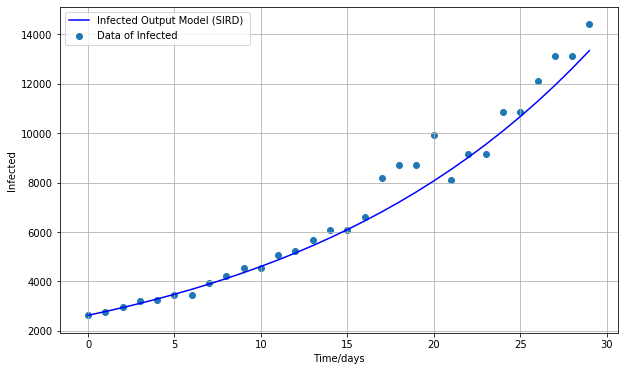}
       \caption{Infected}
       \label{fig:a}
   \end{subfigure}   
   \begin{subfigure}{.5\textwidth}
       \centering
       \includegraphics[scale=0.3]{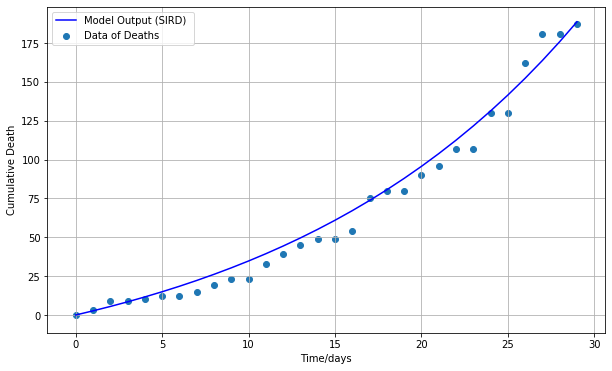}
       \caption{Death}
       \label{fig:b}
   \end{subfigure}
   \caption{The fitted value using the optimal parameters $\beta=1.3201448967113115e-08$ and $~\delta=0.098575$}
   \label{EST}
\end{figure}

The root mean square error (RMSE) \cite{Haug} is a frequently used method to measure  the difference between the values predicted by a model and the values observed in reality. Let $X_{obs}$ be the vector of the observed values and $X_{model}$ the vector of modeled ones. The RMSE of a prediction model with respect to the estimated variable $X_{model}$ is defined as follows
$$RMSE=\sqrt{\frac{1}{n} \sum_{j=1}^{n}\left(X_{model,j}-X_{obs,j}\right)^{2}}$$
Hence, to obtain optimal parameters $\{ \beta, ~\delta \}$ for our model, one should solve the following problem :
$$ min~RMSE $$
Here, the fit is measured by computing the value of the RMSE function using  data of deaths for the beginning of the second wave in Tunisia which is calibrated from September 2021, provided by \footnote{https://covid19.who.int/WHO-COVID-19-global-data.csv} as $X_{obs}$ data.  $X_{model}$ is the death data obtained by the SIRD model (\ref{2.1}) subject to the following initial condition  $$( S(0) = 11718548; ~I(0) = 2629;~ R(0) = 0;~ D(0) = 0)$$
In addition, to minimize the RMSE function, we used the genetic method \footnote{https://github.com/rmsolgi/geneticalgorithm} to update the parameters $\beta$ and $\delta$.  Figure \ref{EST} shows the result of the fitted values using the optimal parameters $\beta$ and $~\delta$.

\section{Conclusion}
In this work, a study of COVID-19 transmission for the case of Tunisia was carried out. A four compartmental mathematical model with mask wear and screening as time-dependent control measures  is developed. The model is proven to have an invariant region where it is well-posed and makes biological sense to be studied for human population. Different properties of the model including global stability analysis of the disease-free and endemic equilibrium points have been studied. Some of the parameter estimates were taken from literature and the remaining parameters were computed  based on real daily data of COVID-19 confirmed cases of Tunisia. The basic reproduction number was also checked.
An optimal analysis of the model for the purpose of assessing the effect of  screening companions was conducted. The result showed that the optimal practice of combination of these two strategies significantly reduces the number of  infections and deaths(see figure \ref{fig:7} ).
In fact, the usage of any form of prevention strategy such as personal mask wear alone led to a decrease in the number of cases in the infected and dead compartments.
The screening applied alone also contributed in lowering the number of infections by lowering the basic reproductive number.
It is also found that combining control strategies mask wear and screening is even better at combating the deadly COVID-19 pandemic in Tunisia .
The optimal application of the control measures (mask wear and screening) though  led to much better and faster results. However, for quicker results, governments are required to set higher maximal values of screening and mask wear (see figure \ref{fig:7}). 
\appendix

\section*{Acknowledgment}
This work was supported in part by the French Ministry for Europe and Foreign Affairs via the project “REPAIR Covid-19-Africa” coordinated by the Pasteur International  Network association  and by European Union's Horizon 2020 research and innovation program under grant agreement No. 883441 (STAMINA).
 \bibliographystyle{elsarticle-num} 
 \bibliography{cas-refs}





\end{document}